# Open and Free Cluster for Public

Z. Akbar[1], Slamet[2], B. I. Ajinagoro[3], G. J. Ohara[3], I. Firmansyah[1], B. Hermanto[1] and L. T. Handoko[1]

[1] *Group for Theoretical and Computational Physics, Research Center for Physics,*
*Indonesian Institute of Sciences, Komplek PUSPIPTEK Serpong, Tangerang 15310, Indonesia*
[2] *PT Prisma Vista Solusi, Jl. MT. Haryono Kav. 15, Jakarta 12810, Indonesia*
[3] *Faculty of Electronics Engineering, STT Telkom, Jl. Telekomunikasi, Dayeuh Kolot,*
*Bandung 40257, Indonesia*

*Abstract*—We introduce the LIPI Public Cluster, the first parallel machine facility fully open for public and for free in Indonesia and surrounding countries. in this paper, we focus on explaining our globally new concept on open cluster, and how to realize and manage it to meet the users needs. We show that after 2 years trial running and several upgradings, the Public Cluster performs well and is able to fulfill all requirements as expected.

*Index Terms*—open cluster, public access, parallel machine.

## I. Introduction

Cluster computer is an extended name of previously known parallel computer or parallel machine. The cluster consists of several single computers at the PC level that are connected each other using usually fast local area network [1]. Utilizing particular softwares mostly as the middleware, it can be viewed in many senses as a single huge computer to accomplish advanced computing which requires high performance hardwares. Previously this requirement was overcomed by so called mainframe computer. However, mainframe is clearly costly and in some senses is lack of flexibility. Especially following the Moore Law on the computer specifications and its prices, the cost performance of any mainframes cannot be matched by mass-produced PC. Nowadays, almost no mainframe has remained in the top 500 of supercomputers in the world [2].

This kind of advanced computings are getting familiar in recent decades in many branches of science, and also in some business areas as well like financial industry. Since there are many types of computation with its own natures and characteristics, a cluster computer should be designed according to the aim of particular computing works which are planned to be done on it. Roughly there are two categories in this issue, processor consuming that is a typical of pure numerical works and space consuming that is a typical of image or database processing [3].

The requirements of cost performance and high specification of computing powers lead to the development of cluster machines. The cost is much lower than constructing a dedicated mainframe, while the performance might be higher depending on the specifications of computer nodes and its network connection. Another point is cluster machine is highly flexible and requires low maintenance cost. The flexibility is realized by its modularity, even one can build a cluster consisting of many types of computers with very different specifications.

Unfortunately in contrast with these facts, starting up a cluster is in most cases not as simple as expected, and of course it is not as cheap as most personal users nor small groups can maintain. This is the main reason for the small number of experts in parallel programmings. In Indonesia especially people working and using parallel programming is very rare, even in academic institutions. We have analyzed that this phenomena is due to the lack of infrastructure, that is the cluster computer itself. Although there are few groups run cluster computers, most of them are not dedicated ones. That means the facility cannot be used in full running for 24 hours a day and 7 days a week. This condition then leads to the result of small number of people having real skills in parallel programmings, that finally discourages any activities requiring parallel programmings too.

We provide a real solution for this problem by introducing the Public Cluster that will be explained in the subsequent section [4]. Before going to summarize the paper we also present briefly its current architecture and some technical details.

## II. The Concept of Public Cluster

In order to overcome the problem in the preceding section, we started developing a cluster facility that is accessible for everyone for free. At the first step of design we assume that our potential users would be the beginners in parallel programmings and who use it for educational or self-learning purposes. Although the cluster is going to be used by our group and some experts collaborating with us, we do not expect any anonymous users with serious computational works. The reason is mainly because the resource is still limited, i.e. number of nodes and its specifications, and we prefer as much as people use our cluster to learn the parallel programmings. In average we plan to provide only 2-3 nodes in a relatively short period, namely less than 3 days, for each anonymous user [5].

These characteristics completely differ with another existing cluster around the world where the clusters are always used by certain people or groups bringing similar type of computational works which are appropriate for the cluster after particular procedures like submitting proposals and any kinds of letter of agreements. In our case, incoming users are completely anonymous and then there are no limitations nor any way to know the type of computational works being executed in the cluster. This also brings another problem to allow job executions owned by different users at same period simultaneously.

Table 1. Generic architecture of cluster [6]

| User's Parallel Applications | | | Fortran/C/C++ Codes | | |
|---|---|---|---|---|---|
| Parallel Environment | | | MPICH/LAM-MPI/OpenMPI | | |
| Software Tools for Applications | | | BLAS/LAPACK/ATLAS/FFT Libraries | | |
| Resource Management Software | | | OpenPBS/TORQUE/MAUI | | |
| System Management Software | | | SSH/C3/SNMP/Ganglia | | |
| OS + Services | Network | Storage | Linux | Ethernet Myrinet Infiniband | NFS/SAN |

Under these assumptions and conditions, the public cluster should fulfill very challenging requirements that are not exist in any conventional clusters:

- Security :
  This is the main issue we should encounter from the first. The user should have enough privilege and freedom to optimize their opportunity to learn parallel programming, while the access must be limited at the maximum level for the sake of security of the whole system and another active user at the same period.
- Flexibility :
  It is impossible to provide the same level of flexibility for anonymous users as well-defined users with direct accesses through ssh etc, but we should allow as much as possible the users to execute their programmes on the cluster. Also there should be a freedom on assigning the number of nodes for each user, since each user could require variously depending on the computational capacities they need.
- Stability :
  Simultaneous executions by different users with various programmes in different blocks of cluster without any interferences among them requires a new innovation on cluster management. This problem includes some techniques to incorporate wide range of nodes with different specifications, that is ranging from 486 to the latest Athlon based nodes.
- Efficiency :
  Since the cluster are dynamically divided into several blocks with various number of nodes inside according to users requests, the nodes should be able to completely turned on or off partially without any interruptions to another working nodes.

Concerning all requirements above, we have then developed the public cluster utilizing some existing tools and self-developed softwares and hardwares to meet the demands mentioned above.

III. THE ARCHITECTURE OF PUBLIC CLUSTER

In principle we follow the general prescription to transform several computers to be a cluster. It requires combination of softwares and hardwares as described in the left column in Table 1. The right column shows the softwares typically used in many clusters. As seen in Table 1, there are several layers to make all nodes in a cluster work together as a single huge computer [6].

However, in our case we should modify this generic architecture to meet particular requirements mentioned before. First, we allow accessing the cluster through web instead of conventional direct access like ssh. The web interface has another advantage of its simplicity and user-friendly that is crucial for less-experienced users. Also it broadens the accessibility since some ISP's or institution do not allow the ssh access for security reason from their networks.

Secondly, we prohibit direct access by the users to the nodes of block provided to them. Instead of that all accesses should go over web through the gateway node. It should be emphasized that the gateway node is not always a master node of their block. In order to realize this management system we have developed a new approach, so called Multi-block Approach [7].

In the middleware, conventional clusters make use of Resource-Management component to allocate the cluster resources according to user request. In public cluster the users are anonymous and then no way to know the computational works and its requirements. On the other hand we have to allocate different block with different number of nodes to each user. Therefore we have deployed our own algorithm to help the administrator to choose appropriately the number and type of nodes among available nodes at present time according to the incoming request. For this purpose we have developed a new algorithm, namely the Extended Genetic Algorithm (EGA) which enables automatic decision making for resource allocation of various nodes for various requirements of applications to obtain the most efficient and optimized environment at a certain period [8]. The EGA could avoid wasting resources and then improve the whole efficiency.

Related to the EGA, we have implemented an integrated and dedicated control and monitoring system based on microcontroller which complements the EGA and web-interface in the aspect of hardware [9]. The system at time being could handle the power supplies, external temperatures and humidities of up to 45 physical nodes. As implemented in the regular big scale clusters, the system also plays an important role in automatic monitoring and control for the working nodes to avoid overheat, to catch up the hardware failures before it happens, etc.

All of these modifications have been well integrated in the web-based interface, either for the end-users or administrators [10]. The rough block diagram describing these modifications is depicted in Fig. 1. As shown in the figure, it is clear that some parts in conventional architecture are discarded, while another new components are inserted.

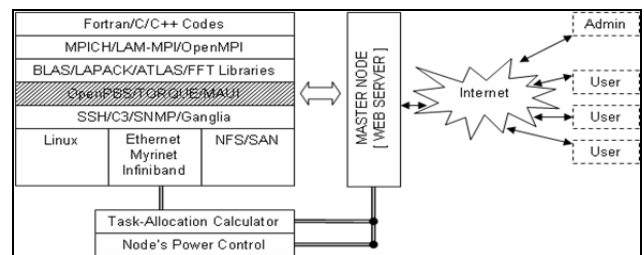

Fig. 1. The architecture of Public Cluster.

## IV. SUMMARY

We have presented the current architecture of LIPI Public Cluster. The cluster is almost totally different with any existing conventional clusters around the world. Its uniqueness however requires some breakthroughs and innovations to realize its main purpose to serve public with an open and free cluster. In fact the usage for public is by definition limited, at least at time being, mostly for educational and self-learning purposes. However this restriction is moreless caused by the limited hardwares we have deployed so far. Technically, the restriction would disappear as the hardwares, the number and its specifications, are improved.

More importantly, such challenging purposes lead to many new inventions in wide area, namely the developments on control and monitoring system, SNMP agent for hardware controls, algorithm for fitness and optimization, unique web-based interface, real-time web-based monitoring and resource allocation management.

Further study is focused on the theoretical aspects like traffic analysis for specific workflow in the current architecture, and the development of middleware customized for this kind of cluster as well.

## V. ACKNOWLEDGMENT

Slamet greatly appreciates valuable discussions with his colleagues at the Faculty of Computer Science, University of Indonesia. B.I. Ajinagoro and G.J. Ohara appreciate their colleagues at the Faculty of Engineering, STT Telkom. This work is financially supported by the Riset Kompetitif LIPI in fiscal year 2007 under Contract no. 11.04/SK/KPPI/II/2007 and the Indonesia Toray Science Foundation Research Grant 2007.